\documentclass[a4paper,10pt,reqno]{amsart}
\usepackage{a4wide}
\usepackage{amsfonts,amsmath,amssymb}
\usepackage{graphicx}
\usepackage{tikz}

\theoremstyle{remark}

\renewcommand{\Re}{\operatorname{Re}}

\def\be{\begin{equation}}
\def\ee{\end{equation}}
\def\bea{\begin{eqnarray}}
\def\eea{\end{eqnarray}}

\begin{document}

\thispagestyle{plain}

\title{On  Thouless  bandwidth formula in the Hofstadter model}

\author{St\'ephane Ouvry (*)  and  Shuang Wu (*)}

\date{\today}

\begin{abstract}
We generalize Thouless  bandwidth formula to its $n$-th moment. We obtain a closed expression in terms of polygamma, zeta and Euler numbers. 
\end{abstract}

\maketitle

(*) LPTMS, CNRS-Facult\'e des Sciences d'Orsay, Universit\'e Paris Sud, 91405 Orsay Cedex, France

\section{ Introduction }

In a series of stunning papers stretching over almost a  decade \cite{Thouless}  Thouless  obtained a closed expression for the bandwidth of the Hofstadter spectrum \cite{Hofstadter} in the $q\to\infty$  limit. 
Here the integer $q$  stands for the denominator of the rational flux $
\gamma  = 2 \pi {p}/{q}
$ of the  magnetic field piercing a unit cell of the square lattice; the numerator $p$ is taken to be $1$ (or equivalently $q-1$)\footnote{In the sequel $p$ is  understood to be equal to $1$.}.

Let us recall that in the commensurate case  where the lattice eigenstates $\psi_{m,n}= e^{i n k_y}\Phi_m$ are $q$-periodic $\Phi_{m + q} = e^{i q k_x } \Phi_m$, with   $k_x, k_y\in [-\pi,\pi]$, 
  the Schrodinger   equation 
\be\Phi_{m+1}+\Phi_{m-1}+2\cos(k_y+\gamma m)\Phi_{m}=e\Phi_{m}\label{eq}\ee 
reduces  
 to the $q\times q$ secular matrix
\be\nonumber m_{p/q}(e,k_x,k_y)=\begin{pmatrix}
 2 \cos ({k_y})-e& 1 & 0 & \cdots & 0 & e^{-i {q k_x}} \\
1 & 2 \cos ({k_y}+\frac{2\pi p}{q})-e & 1 & \cdots & 0 & 0 \\
0 & 1 & () & \cdots & 0 & 0 \\
\vdots & \vdots & \vdots & \ddots & \vdots & \vdots \\
0 & 0 & 0  & \cdots & () & 1 \\
e^{i {q k_x}}  & 0 & 0  & \cdots & 1 & 2 \cos ({k_y}+(q-1)\frac{2\pi p}{q})-e  \\
\end{pmatrix}  \ee
acting as 
\be\label{sharp} m_{p/q}(e,k_x,k_y).\Phi=0\ee 
on the $q$-dimensional eigenvector $\Phi=\{\Phi_{0}, \Phi_{1},\ldots, \Phi_{q-1}\}$.
Thanks to the identity  
\be\nonumber \det(m_{p/q}(e,k_x,k_y))=\det(m_{p/q}(e,0,0))-2 (-1)^q (\cos(q k_x)-1 + \cos(q k_y)-1),\ee
 the  Schrodinger equation   $\det(m_{p/q}(e,k_x,k_y))=0$  rewrites \cite{Chambers} as
\be\label{eigen} \det(m_{p/q}(e,0,0))=2 (-1)^q (\cos(q k_x)-1 + \cos(q k_y)-1) \ee
The polynomial  
\be\nonumber b_{p/q}(e)=-\sum_{j=0}^{[{q\over 2}]}a_{p/q}(2j)e^{2j}\ee 
materializes in  $\det(m_{p/q}(e,0,0))$  
\be\label{sososimple}\det(m_{p/q}(e,0,0))+4(-1)^q=(-1)^q e^qb_{p/q}(1/e)
\ee
so that
eq.~(\ref{eigen}) becomes
\begin{equation}\nonumber
e^qb_{p/q}(1/e)=2(\cos(q k_x)+\cos(q k_y)).
\end{equation} 
The $a_{p/ q}(2j)$'s (with $a_{p/q}(0)=-1$)  are  related  to the Kreft  coefficients  \cite{Kreft}: 
how to  get an explicit expression for  these coefficients 
is explained in Kreft's paper.

We focus on the Hofstadter spectrum bandwidth  defined in terms of 
 the $2q$ edge-band energies $e_r(4)$ and $e_r(-4)$, $r=1,2,\ldots, q,$ solutions    of
\be\nonumber e^q b_{p/ q}(1/e)= 4\;{\rm and}\;e^q b_{p/ q}(1/e)=- 4\ee 
respectively (see Figures 1 and 2).
\begin{figure}
\includegraphics[scale=.7]{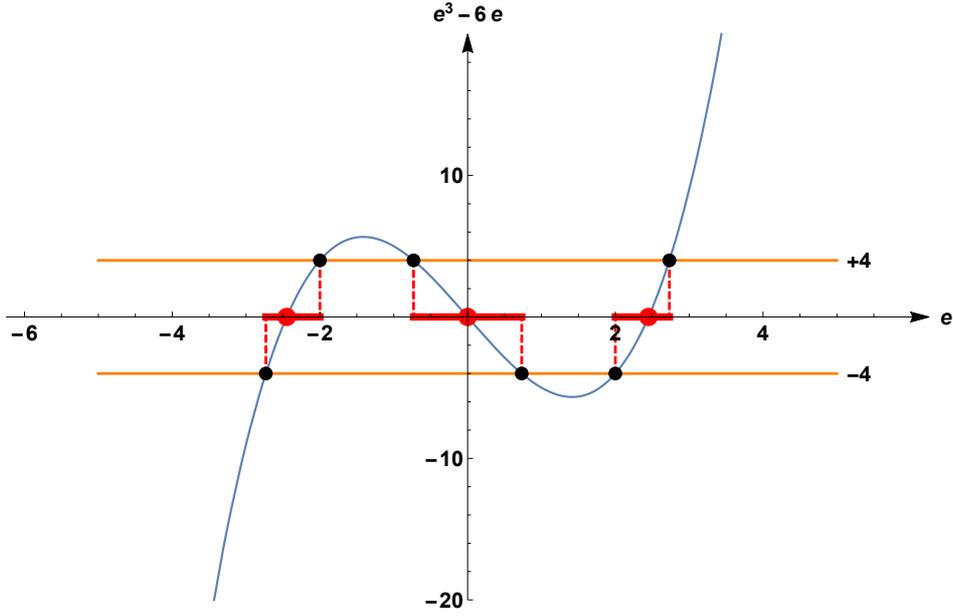}\label{figure1}
\caption{$p=1, q=3, e^q b_{p/ q}(1/e)=e^3-6e:$  the 3 horizontal red segments are the energy bands; the 3 red dots are the mid-band energies; the 6 black dots are the $\pm 4$ edge-band energies.}
\end{figure}
  \begin{figure}
\includegraphics[scale=.7]{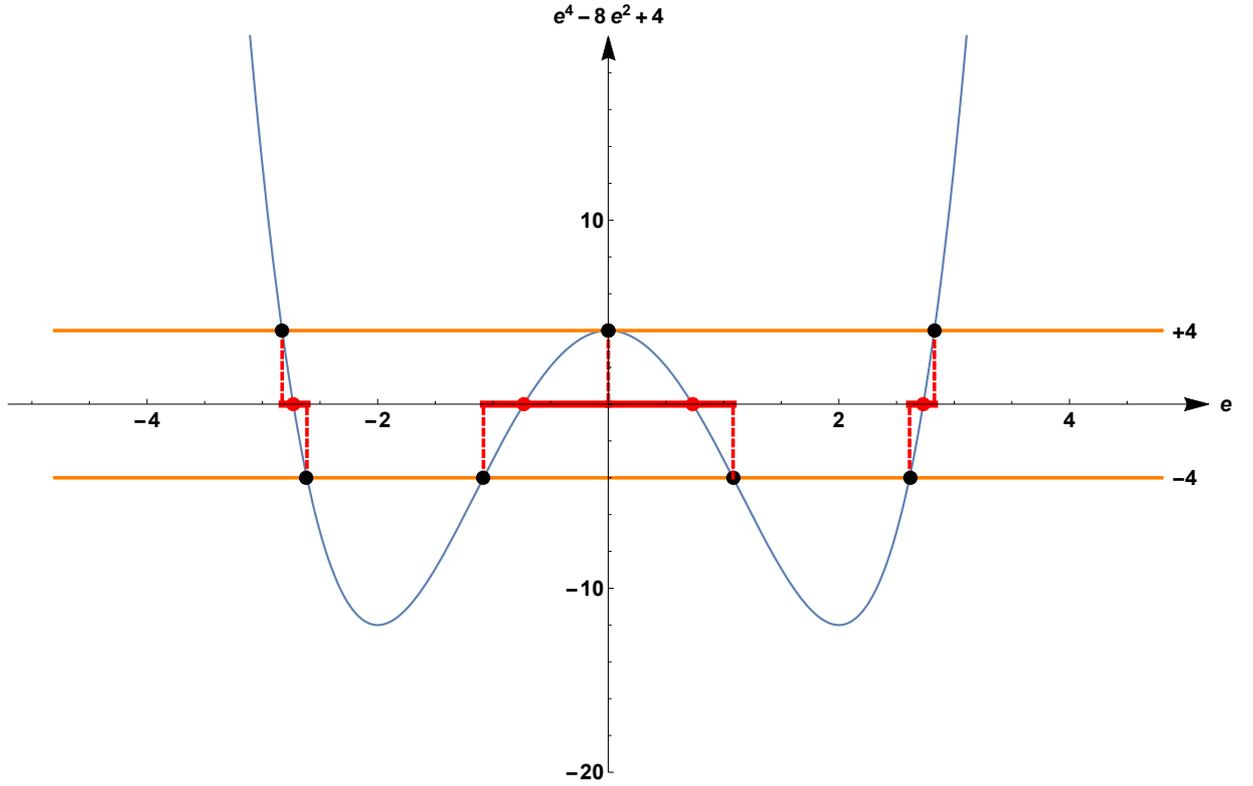}\label{figure1bis}
\caption{$p=1, q=4, e^q b_{p/ q}(1/e)=e^4-8e^2+4:$  the 4 horizontal red segments are the energy bands; the 4 red dots are the mid-band energies; the 8 black dots are the $\pm 4$ edge-band energies --there are two degenerate dots located at the center.}
\end{figure}
If one specifies an ordering  for the $e_r(4)$'s and the $e_r(-4)$'s 
\be\nonumber e_1(4)\le e_2(4)\le \ldots\le e_q(4)\quad{\rm and}\quad  e_1(-4)\le e_2(-4)\le \ldots\le e_q(-4)\ee
the   bandwidth  is 
\be \label{band} (-1)^{q+1}\sum_{r=1}^q (-1)^r \big(e_r(-4)-e_r(4)\big).\ee 
The  Thouless formula is obtained  in the $q\to\infty$  limit as
\be\label{Thouless} \lim_{q\to\infty} (-1)^{q+1} q \sum_{r=1}^q (-1)^r \big(e_r(-4)-e_r(4)\big)={32\over \pi}\sum_{k=0}^\infty(-1)^k{1\over (2k+1)^2}\ee
(see also \cite{Last}).
We aim to extend this result to the  
 $n$-th moment  defined as
\be \label{nmoment}  (-1)^{q+1}\sum_{r=1}^q (-1)^r \big(e_r^n(-4)-e_r^n(4)\big).\ee 
which  is  a natural generalization\footnote{Computing also the bandwidth $n$-th moment 
\be\label{graal}  (-1)^{q+1}\sum_{r=1}^q (-1)^r \big(e_r(-4)-e_r(4)\big)^n,\ee here defined for $n$ odd, would be of particular interest. We will come back to this question in the conclusion.} of (\ref{band}): one can think of it  as 
\be \nonumber{ n}\int_{-4}^4 \tilde{\rho}_{p/q}^{}(e)e^{n-1}de\ee
where $\tilde{\rho}_{p/q}^{}(e)$ is the indicator function with value 1 when $|e^qb_{p/q}(1/e)|\le 4$  and 0 otherwise.
 
  Trivially (\ref{nmoment})  vanish when $n$ is even --we will see later  how to give   a non trivial meaning to the $n$-th moment in this case.  
 Therefore we focus on (\ref{nmoment})  when  $n$ is odd 
and, additionnally, when $q$ is odd, in which case it simplifies further to 
\be \label{hard} -2\sum_{r=1}^q (-1)^r e_r^n(4) = 2\sum_{r=1}^q (-1)^r e_r^n(-4)\ee
thanks to the symmetry $e_{r}(-4)=-e_{q+1-r}(4)$.  
As said above, the $e_r(4)$'s are the roots of $e^q b_{p/ q}(1/e)= 4$  that is, by the virtue of (\ref{sososimple}), those of 
\be\nonumber \det(m_{p/q}(e,0,0))=0.\ee
\section{The first moment : Thouless formula}
The key point in the observation of Thouless \cite{Thouless}  is that if evaluating the first moment rewritten in (\ref{hard}) as $2\sum_{r=1}^q (-1)^r e_r(4)$ when $q$ is odd seems at first sight untractable, still,  
\begin{itemize}
\item thanks to $\det(m_{p/q}(e,0,0))$  factorizing as
\be \nonumber\det(m_{p/q}(e,0,0))=-\det(m_{p/q}^{++}(e))\det(m_{p/q}^{--}(e))\ee
where 
\be \nonumber m_{p/q}^{++}(e)=\begin{pmatrix}
 e-2 & 2 & 0 & \cdots & 0 & 1 \\
1 & e-2 \cos (\frac{2\pi p}{q}) & 1 & \cdots & 0 & 0 \\
0 & 1 & () & \cdots & 0 & 0 \\
\vdots & \vdots & \vdots & \ddots & \vdots & \vdots \\
0 & 0 & 0  & \cdots & () & 1 \\
1  & 0 & 0  & \cdots & 1 & e-2 \cos ({q-1\over 2}\frac{2\pi p}{q})-1  \\
\end{pmatrix}  \ee
\be \nonumber m_{p/q}^{--}(e)=\begin{pmatrix}
 e-2\cos (\frac{2\pi p}{q}) &1 & 0 & \cdots & 0 & 1 \\
1 &e- 2 \cos (\frac{4\pi p}{q}) & 1 & \cdots & 0 & 0 \\
0 & 1 & () & \cdots & 0 & 0 \\
\vdots & \vdots & \vdots & \ddots & \vdots & \vdots \\
0 & 0 & 0  & \cdots & () & 1 \\
1  & 0 & 0  & \cdots & 1 & e-2 \cos ({q-1\over 2}\frac{2\pi p}{q})+1  \\
\end{pmatrix}  \ee
are matrices of size $(q+1)/2$ and $(q-1)/2$ respectively, so that the $e_r(4)$'s  split in two packets  $e_r^{++}$, $r=1,2,\ldots, (q+1)/2,$ the roots of $\det(m_{p/q}^{++}(e))=0$  and $e_r^{--}$, $r=1,2,\ldots, (q-1)/2,$   those of $\det(m_{p/q}^{--}(e))=0$
 \item and thanks to $\sum_{r=1}^q (-1)^r e_r(4)$ happening to rewrite as 
\be\nonumber -\sum_{r=1}^q (-1)^r e_r(4)=\sum_{r=1}^{q+1\over 2}  |e_r^{++}|-\sum_{r=1}^{q-1\over 2}  |e_r^{--}|\ee  
\end{itemize}
 (\ref{hard}) becomes tractable since it reduces to the sum of the absolute values of the roots of two  polynomial equations.

Indeed using \cite{Thouless} 
\be \nonumber {2i\over \pi}\int_{-i x}^{i x} \left({z\over z-a}-1\right)dz={4a\over\pi} \arctan({x\over a}),\ee
\be\nonumber\lim_{x\to\infty}{4a\over\pi} \arctan({x\over a})=2|a|\ee
and
 \be \nonumber \lim_{x\to\infty}{2i\over\pi }\int_{-i x}^{i x} \left({z\over z-a}-1\right) dz=\lim_{x\to\infty}{2i\over \pi} \int_{-i x}^{i x}\left(-\log{z - a\over z}\right)dz\ee 
one gets  
\be \label{ratio}2\left(\sum_{r=1}^{q+1\over 2}  |e_r^{++}|-\sum_{r=1}^{q-1\over 2}  |e_r^{--}|\right)={2i\over \pi}\lim_{x\to\infty}\int _{-i x}^{i x}\log\left({ z \det(m_{p/q}^{--}(z))\over \det(m_{p/q}^{++}(z))}\right).\ee
Making \cite{Thouless}  further algebraic manipulations on the ratio of determinants in (\ref{ratio}) in  particular in terms of particular solutions $\{\Phi_{0}, \Phi_{1},\ldots, \Phi_{q-1}\}$ of (\ref{sharp}) --on the one hand $\Phi_0=0$ and on the other hand $\Phi_{(q-1)/2}=\Phi_{(q+1)/2}$--  and then for large $q$  taking in (\ref{eq}) the continuous limit  lead to, via the change of variable $y=qz/(8\pi i)$,   
\be \nonumber \lim_{q\to\infty} 2q\left(\sum_{r=1}^{q+1\over 2}  |e_r^{++}|-\sum_{r=1}^{q-1\over 2}  |e_r^{--}|\right)={32}\int_0^{\infty}\log\left({\Gamma(3/4+y)^2\over y\Gamma(1/4+y)^2}\right)dy\ee
This last integral    gives   the first moment  
\be \label{Thoulessbis}\lim_{q\to\infty} q\left(\sum_{r=1}^q (-1)^r \big(e_r(-4)-e_r(4)\big)\right)={4\over \pi}\left(\psi ^{(1)}\left(\frac{1}{4}\right)-\pi^2\right)\ee which is a rewriting of
(\ref{Thouless})  ($\psi ^{(1)}$ is the polygamma function of order 1). 
\section{The $n$-th moment }
\subsection{$n$ odd:}
to evaluate  the $n$-th moment one follows the steps above by first noticing that
\be \nonumber -\sum_{r=1}^q (-1)^r e_r^n(4)=\sum_{r=1}^{q+1\over 2}  |e_r^{++}|^n-\sum_{r=1}^{q-1\over 2}  |e_r^{--}|^n\ee  
 holds.
 Then using 
 \be\nonumber  {2i\over \pi} \int_{-i x}^{i x} \left({z^n\over z-a}-\sum_{k=0}^{n-1} a^k z^{n-1-k} \right)dz={{4a^n\over\pi} \arctan({x\over a})},\ee
 \be\nonumber \lim_{x\to\infty}{{4a^n\over\pi} \arctan({x\over a})} =2|a^n|\ee 
 and
 \be \label{toto}\lim_{x\to\infty}{2i\over\pi }\int_{-i x}^{i x} \left({z^n\over z-a}-\sum_{k=0}^{n-1} a^k z^{n-1-k} \right)dz=\lim_{x\to\infty}{2i\over \pi} \int_{-i x}^{i x}-nz^{n-1}\left(\log{z - a\over z}+\sum_{k=1}^{n-1}{a^k\over k z^k} \right)dz\ee
 one gets 
\be \label{hardbis}2\left(\sum_{r=1}^{q+1\over  2}  |e_r^{++}|^n-\sum_{r=1}^{q-1\over 2}  |e_r^{--}|^n\right)={2i\over \pi}\lim_{x\to\infty}\int _{-i x}^{i x}nz^{n-1}\left(\log\left({ z \det(m_{p/q}^{--}(z))\over \det(m_{p/q}^{++}(z))}\right)-\sum_{k=1}^{n-1}{\sum_{r=1}^{q+1\over 2}(e_r^{++})^k-\sum_{r=1}^{q-1\over 2}(e_r^{--})^k\over k z^k}\right)dz.\ee
 In the RHS of (\ref{toto}) the polynomial  $z^{n-1}\sum_{k=1}^{n-1}{a^k\over k z^k}$   cancels 
the positive or nul exponents in the expansion   around $z=\infty$ of the logarithm term $z^{n-1}\log{z - a\over z}$. Likewise, in (\ref{hardbis}), the same mechanism  takes  place   for $-z^{n-1}\left(\sum_{k=1}^{n-1}{\sum_{r=1}^{q+1\over 2}(e_r^{++})^k-\sum_{r=1}^{q-1\over 2}(e_r^{--})^k\over k z^k}\right)$  with respect to   $z^{n-1}\log\left({ z \det(m_{p/q}^{--}(z))\over \det(m_{p/q}^{++}(z))}\right)$.
Additionally,    the polynomials 
can be reduced to their  $k$ even components. 
Further algebraic manipulations in (\ref{hardbis})  and,  when  $q$ is large,
 taking the continous limit,   lead to,  via the change of variable $y=qz/(8\pi i)$, 
\be \label{finalbis}\lim_{q\to\infty}2 q^n\left(\sum_{r=1}^{q+1\over 2}  |e_r^{++}|^n-\sum_{r=1}^{q-1\over 2}  |e_r^{--}|^n\right)={(8\pi i)^{n-1}32}\int_0^{\infty}n y^{n-1}\left(\log\left({\Gamma(3/4+y)^2\over y\Gamma(1/4+y)^2}\right)+\sum_{k=2, k\;{\rm even}}^{n-1}{E_k\over k 4^k y^k} \right)dy. \ee
To go from (\ref{hardbis}) to (\ref{finalbis}) one has  used that for  $k$  even, necessarily\footnote{(\ref{infinity})  is also strongly supported  by numerical simulations. More generally the $k$-th moments $\sum_{r=1}^{q+1\over 2}(e_r^{++})^k$ and $\sum_{r=1}^{q-1\over 2}(e_r^{--})^k$  can  be directly  retrieved  from the coefficients of  $\det(m_{p/q}^{++}(e))$ and $\det(m_{p/q}^{--}(e))$ respectively. In particular one finds  $\sum_{r=1}^{q+1\over 2}e_r^{++}=2$ and $\sum_{r=1}^{q-1\over 2}e_r^{--}=-2$;  for  $k$  odd  $\lim_{q\to\infty}\sum_{r=1}^{q+1\over 2}(e_r^{++})^k=4^k/2$ and $\lim_{q\to\infty}\sum_{r=1}^{q-1\over 2}(e_r^{--})^k=-4^k/2$; for  $k$ even  $\lim_{q\to\infty}1/q \left(\sum_{r=1}^{q-1\over 2}(e_r^{++})^k\right)=\lim_{q\to\infty}1/q\left(\sum_{r=1}^{q-1\over 2}(e_r^{--})^k\right) =\binom{k}{{k}/{2}}^2/2$. This last result can easily be understood in terms of the number $\binom{k}{{k}/{2}}^2$ of closed lattice  walks  with $k$ steps \cite{nous}.}  
   \be \label{infinity} \lim_{{q\to\infty}}q^k\left(\sum_{r=1}^{q+1\over 2}(e_r^{++})^k-\sum_{r=1}^{q-1\over 2}(e_r^{--})^k\right)=(2\pi)^k |E_k| \ee  where the $E_k$'s  are the Euler numbers. 
 Indeed in (\ref{finalbis}), as it was the case in   (\ref{toto},\ref{hardbis}),  the polynomial $\sum_{k=2,k\;{\rm even}}^{n-1}E_k/(k 4^k) y^{n-1-k}$   cancels the positive or nul exponents in the expansion  around $y=\infty$ of the logarithm term $y^{n-1}\log\left({\Gamma(3/4+y)^2\over y\Gamma(1/4+y)^2}\right)$ \cite{Wagner}. 
 It amounts to a fine tuning at the  infinite upper integration limit so that after integration the end result   is finite.   
Performing this last integral  gives   the $n$-th moment 
\be \label{final}\lim_{q\to\infty} q^n\left( \sum_{r=1}^q (-1)^r \left(e_r^n(-4)-e_r^n(4)\right)\right)={4\over \pi }\left((-1)^{n-1}\psi ^{(n)}\left(\frac{1}{4}\right)-2^n
   \left(2^{n+1}-1\right) \zeta (n+1) n!\right)\ee
which generalizes the Thouless formula (\ref{Thoulessbis}) to  $n$ odd ($\psi ^{(n)}$ is the polygamma function of order $n$).

\subsection{$n$ even:}
as said above  the   $n$-th moment trivially  vanishes when $n$ is even. In this case, we should rather consider a $n$-th moment  restricted to the positive --or equivalently  by symmetry  negative--  half of the spectrum\footnote{Instead of ${ n}\int_{-4}^4 \tilde{\rho}_{p/q}^{}(e)e^{n-1}de$ one considers \be\nonumber{ n}\int_{-4}^0 \tilde{\rho}_{p/q}^{}(e)e^{n-1}de={ n}\int_{0}^4 \tilde{\rho}_{p/q}^{}(e)e^{n-1}de.\ee.}. In the $q$  odd case 
 it is 
   \be\label{half}   -\sum_{r=1}^{(q-1)/2} (-1)^r\left(e_r^n(-4)- e_r^n(4)\right)+e_{(q+1)/ 2}^n\left((-1)^{q+1\over 2}4\right)= \sum_{(q+3)/2}^{q} (-1)^r\left(e_r^n(-4)- e_r^n(4)\right)+e_{(q+1)/2}^n\left((-1)^{q-1\over 2}4\right).\ee
   It is still true that
   \be \nonumber \sum_{r=1}^{q+1\over 2}  (e_r^{++})^n-\sum_{r=1}^{q-1\over 2}  (e_r^{--})^n=\sum_{(q+3)/2}^{q} (-1)^r\left(e_r^n(-4)- e_r^n(4)\right)+e_{(q+1)/2}^n\left((-1)^{q-1\over 2}4\right)\ee
 where, since $n$ is even,  absolute values are not needed anymore, a simpler situation.
It follows that the right hand side of  (\ref{final}) also gives, when $n$ is even,   
   twice the  $q\to\infty$ limit of
     the   half  spectrum $n$-th moment  as defined in (\ref{half}), up to a factor $q^n$.
 
 \subsection{Any $n$:}  one reaches the conclusion that  \bea\label{simple}&&{4\over \pi }\left((-1)^{n-1}\psi ^{(n)}\left(\frac{1}{4}\right)-2^n
   \left(2^{n+1}-1\right) \zeta (n+1) n!\right)\\ \nonumber&=&\frac{2}{\pi } 4^{n+1} n! \sum _{k=0}^{\infty } \frac{(-1)^k}{(2
   k+1)^{n+1}}\\ \nonumber &=&\frac{2}{\pi } n! \big(\zeta (n+1,\frac{1}{4})-\zeta
   (n+1,\frac{3}{4})\big)\eea yields $q^n$ times the  $n$-th moment when $n$ is odd\footnote{When $n$ is odd it is also  twice the half spectrum $n$-th moment
   \be\nonumber { n}\int_{-4}^4 \tilde{\rho}_{p/q}^{}(e)e^{n-1}de=2 { n}\int_{0}^4 \tilde{\rho}_{p/q}^{}(e)e^{n-1}de.\ee.} and twice the half spectrum $n$-th moment when $n$ is even. Numerical simulations do confirm convincingly this result (eventhough the convergence is slow). In the $n$ even case one already knows from (\ref{infinity}) that (\ref{simple}) simplifies further to 
   \be  \nonumber 
    2|E_n|\big({2\pi}\big)^n \ee 
from which one gets for the  $n\to\infty$-moment   scaling
   \be\nonumber  \mu \; n! 2^{2 n} \ee
   where  \be \nonumber \mu=\lim_{n\to\infty, n {\rm\;even}}\frac{2^{1-n} \pi ^n |E_n|}{n!}=2.54647\ldots.\ee

 \section{Conclusion and opened issues} 
  
(\ref{simple}) is certainly a simple and convincing $n$-th moment generalization of the Thouless bandwidth formula (\ref{Thouless}). It remains to be proven on more solid grounds for example in the spirit   of   \cite{Last}. 

In the definition of the $n$-th moment (\ref{nmoment}) one can view the exponent $n$  as a magnifying loop of the Thouless first moment. 
(\ref{simple}) was obtained for $p=1$ (or $q-1$): it would certainly be interesting to understand what happens for $p\ne 1$  where numerical simulations indicate  a strong $p$ dependence  when $n$ increases, an effect of the $n$-zooming inherent to the  $n$-th moment definition  (\ref{nmoment}). 

In the $n$ even case, twice the half spectrum  $n$-th moment ends up being equal to  $2 |E_n|({2\pi/ q})^n $, a result that  can be interpretated  as if, at the $n$-zooming level, they were $2|E_n|$ bands each of length ${2\pi/ q}$. 
It would  be interesting to  see if this Euler counting has  a meaning in the context of lattice walks  \cite{nous} (twice the Euler number $2|E_n|$ counts the number of  alternating permutations in $S_n$).

Finally,  returning to the bandwidth $n$-th moment defined in (\ref{graal})  for $n$ odd, and focusing again on  $q$ odd, one can expand 
\be\label{expansion}\sum_{r=1}^q (-1)^r \big(e_r(-4)-e_r(4)\big)^n=-2\sum_{k=0}^{(n-1)/2}\binom{n}{k}(-1)^k\sum_{r=1}^q (-1)^r e_r(-4)^ke_r(4)^{n-k}\ee  where the symmetry $e_{r}(-4)=-e_{q+1-r}(4)$ has again been used. The $k=0$ term $-2\sum_{r=1}^q (-1)^r e_r(4)^{n}$ is the $n$-th moment discussed above
and one knows that multiplying it by $q^n$ ensures in the $q\to\infty$ limit a finite scaling. Let us also  multiply in  (\ref{expansion})   the   $k=1,\ldots,(n-1)/2$ terms by $q^n$: one  checks numerically that 
\bea\nonumber \lim_{q\to\infty}-2q^n \sum_{r=1}^q (-1)^r e_r(-4)^ke_r(4)^{n-k}&=&{n-2k\over n}\lim_{q\to\infty}-2q^n \sum_{r=1}^q (-1)^r e_r(4)^{n}
\\\nonumber&=&{n-2k\over n}{4\over \pi }\left((-1)^{n-1}\psi ^{(n)}\left(\frac{1}{4}\right)-2^n
   \left(2^{n+1}-1\right) \zeta (n+1) n!\right).\eea 
Using \be\nonumber\sum_{k=0}^{(n-1)/2}\binom{n}{k}(-1)^k{n-2k\over n}=0\ee one  concludes that in the $q\to\infty$ limit 
the  bandwidth $n$-th moment  is  such that \be \nonumber\lim_{q\to\infty}q^n\sum_{r=1}^q (-1)^r \big(e_r(-4)-e_r(4)\big)^n=0\ee  when $n$ is odd, a fact which is also supported by numerical simulations\footnote{Similarly, when   $n$ is even,  the bandwidth $n$-th moment now defined  as \be \nonumber\sum_{r=1}^q  \big(e_r(-4)-e_r(4)\big)^n\ee is such that  $\lim_{q\to\infty} q^n\sum_{r=1}^q  \big(e_r(-4)-e_r(4)\big)^n=0$.}. Clearly, multiplying   the  sum  in (\ref{expansion}) by  $q^n$ is insufficient,  a  possible manifestation of the   fractal structure \cite{Last} of the band spectrum. 
We leave  to further studies the question of finding a right scaling for  the bandwidth $n$-th moment.

\section{Acknowledgements}

S. O. acknowlegdes interesting discussions with Eug\`ene Bogomolny and Stephan Wagner and thank Alain Comtet    for a  careful reading of the manuscript. Discussions with Vincent Pasquier are also acknowledged.

\end{document}